\documentstyle[graphicx,amsfonts,amssymb,amsmath,aps,twocolumn,prd,url]{revtex}

\begin{document}

\bibliographystyle{prsty}



\title{Cherenkov Radiation from $e^+e^-$ Pairs and Its Effect on $\nu_e$ Induced Showers}

\author{Sourav K. Mandal, Spencer R. Klein, J. David Jackson}
\address{Lawrence Berkeley National Laboratory\\Berkeley, CA 94720}


\maketitle

\begin{abstract}

We calculate the Cherenkov radiation from an $e^+e^-$ pair at small
separations, as occurs shortly after a pair conversion.  The radiation
is reduced (compared to that from two independent particles) when the
pair separation is smaller than the wavelength of the emitted light.
We estimate the reduction in light in large electromagnetic showers,
and discuss the implications for detectors that observe Cherenkov
radiation from showers in the Earth's atmosphere, as well as in oceans
and Antarctic ice.

\end{abstract}



\section{Introduction}

Cherenkov radiation from relativistic particles has been known for
over 70 years \cite{firstpaper}.  However, to date, almost all studies
have concentrated on the radiation from individual particles.  Frank
\cite{frank}, Eidman\cite{eidman} and Balazs \cite{balazs} considered
the Cherenkov radiation from electric and magnetic dipoles, but only
in the limit of vanishing separations $d$.  Their work was nicely reviewed
by Jelley \cite {jelley}.
 
Several more recent calculations have considered Cherenkov radiation
from entire electromagnetic showers, in the coherent or almost
coherent limit \cite{zhs}.  The fields from the $e^+$ and $e^-$
largely cancel, and the bulk of the coherent radiation is due to the
net excess of $e^-$ over $e^+$ (the Askaryan effect) \cite{askaryan}.
Hadronic showers produce radiation through the same mechanism
\cite{zhadron}.  Coherent radiation occurs when the wavelength of the
radiation is large compared to the radial extent of the shower; for
real materials, this only occurs for radio waves.
 
Here, we consider another case, the reduction of radiation from
slightly-separated oppositely-charged co-moving pairs.  This includes
$e^+e^-$ pairs produced by photon conversion.  When high-energy
photons convert to $e^+e^-$ pairs, the pair opening angle is small and
the $e^+$ and $e^-$ separate slowly.  

Near the pair, the electric and magnetic fields from the $e^+$ and
$e^-$ must be considered separately.  However, for an observer far
away from the pair (compared to the pair separation $d$), the electric and
magnetic fields from the $e^+$ and $e^-$ largely cancel.  Cherenkov
radiation is produced at a distance of the order of the photon
wavelength $\Lambda$ from the charged particle trajectory.  So, for
$d<\Lambda$, cancellation reduces the Cherenkov radiation from a pair
to below that for two independent particles.  For a typical pair
opening angle $m/k$, where $k$ is the photon energy and $m$ the
electron mass, without multiple scattering, $\Lambda > d$ for a
distance $k\Lambda/m$.  For blue light ($\Lambda = 400$ nm) from a 1
TeV pair, the radiation is reduced until the pair travels a distance
of 40 cm (neglecting multiple scattering).

A similar cancellation effect was observed for energetic ($\sim 100$
GeV) $e^+e^-$ pairs in nuclear emulsions \cite{perkins}.  Ionization
from newly created $e^+e^-$ pairs is reduced when the pair separation
is less than the screening distance for ionization in the target.

In this paper, we calculate the Cherenkov radiation from $e^+e^-$
pairs, simulate optical radiation from pairs follow realistic
trajectories, and consider the radiation from electromagnetic showers.
We consider two classes of experiments: underwater/in-ice neutrino
observatories and air Cherenkov telescopes.

\section{Cherenkov radiation from pairs}

Cherenkov radiation from closely spaced $e^+e^-$ pairs can be derived
by extending the derivation for point charges, by replacing a point
charge with an oppositely charged, separated pair.  We sketch the
derivation for radiation from point charges, review previous work on
radiation from infinitesimal dipoles, and derive the expression for
Cherenkov radiation from a closely-spaced co-moving pair.

We follow the notation and derivation from Ref. \cite{jackson}.  In
Fourier space, the charge density $\rho$ and current density $\vec{J}$ from a
point charge $ze$ propagating with speed $v$ in the $x_1$ direction
can be written as
\begin{equation}
\label{eq:mono current density}
\begin{aligned}
\rho(\vec k, \omega) &= \frac{ze}{2\pi}\delta(\omega-k_1 v)
\\
\vec J(\vec k, \omega) &= \vec v \rho(\vec k, \omega)
\end{aligned}
\end{equation}
where $\vec{k}$ is the wave vector and $\omega$ the photon energy.
This current deposits energy into the medium through electromagnetic
interactions.  We use Maxwell's equations beyond a radius $a$ around
the particle track, where $a$ is comparable to the average atomic
separation.  Then, by conservation of energy, the Cherenkov radiation
power is equal to the the energy flow through a cylinder of this
radius, giving
\begin{equation}
\left(\frac{dE}{dx}\right) = -ca Re\int_0^\infty B_3^\ast(\omega)E_1(\omega)d\omega\;.
\end{equation}
$E_1$ is the component of $\vec E$ parallel to the particle track, and
$B_3$ is the component of $\vec B$ in the $x_3$ direction, evaluated
at an impact parameter $b$ at a point with $x_2=b$, $x_3=0$.  We omit
the time-phase factors for brevity.

Using the wave equations in a dielectric medium and the definition of
fields, then integrating over momenta, which eliminates the space-phase factors, one finds
\begin{equation}
E_1(\omega) = -\frac{ize\omega}{v^2}\left(\frac{2}{\pi}\right)^{1/2}\left[\frac{1}{\epsilon(\omega)}-\beta^2\right]K_0(\lambda b)\label{eq:mono E1}
\end{equation}
where
\[\lambda^2 = \frac{\omega^2}{v^2}[1-\beta^2\epsilon(\omega)]\;.\]
Similarly,
\begin{equation}
\begin{aligned}
E_2(\omega)&=\frac{ze}{v}\left(\frac{2}{\pi}\right)\frac{\lambda}{\epsilon(\omega)}K_1(\lambda b)\\
B_3(\omega) &= \epsilon(\omega)\beta E_2(\omega)\;.
\end{aligned}
\end{equation}
$K_0$ and $K_1$ are the zeroth and first order modified Bessel functions of the second kind. 

The far-field radiation depends on the asymptotic form of the energy
deposition at $|\lambda a|\gg 1$.  For $\beta >
1/\sqrt{\epsilon(\omega)}$ for real $\epsilon(\omega)$, $\lambda$ is
completely imaginary.  The asymptotic contribution of the Bessel
functions in the integrand of $dE/dx$ is finite, giving the well-known
expression for the Cherenkov radiation
\begin{equation}
\begin{aligned}
\left(\frac{dE}{dx}\right) &=\frac{(ze)^2}{c^2}\\
&\times\int_{\epsilon(\omega)>1/\beta^2}
\omega \left(1-\frac{1}{\beta^2\epsilon(\omega)}\right)d\omega\;.
\end{aligned}
\label{eq:mono dE/dx}
\end{equation}
Note how $a$ has dropped out \cite[Ch. 13]{jackson}. The derivation
of this Cherenkov radiation may be expanded to give the field
from a pair. 

The radiation from an $e^+e^-$ pairs depends on two parameters: the
separation $d$ and the angle between the direction of motion and the
orientation of the pair.  For relativistic pairs created by photon
conversion, the transverse (to the direction of motion) separation is
important; the longitudinal separation of a highly relativistic pair
can be neglected, due to Lorentz length contraction.

Balazs \cite{balazs} provided an expression for Cherenkov radiation
from an infinitesimal dipole $D$ oriented transverse to its momentum.
The fields are approximated by by a linear Taylor expansion of
the corresponding point-charge fields:

\begin{align*}
E_1^{(D)}(\omega) = -d \frac{\partial E_1(\omega)}{\partial x_2}\;;\;
B_3^{(D)}(\omega) = -d \frac{\partial B_3(\omega)}{\partial x_2}
\end{align*}
where $d$ is the effective pair separation, so $D=zed$.  Then,
following the same steps as in the point-charge case, Balazs finds
\begin{equation}
\begin{aligned}
\left(\frac{dE}{dx}\right) &=\frac{1}{2}\frac{D^2}{c^4}\\
&\times\int_{\epsilon(\omega)>1/\beta^2}
\epsilon(\omega)\omega^3\left(1-\frac{1}{\beta^2\epsilon(\omega)}\right)^2d\omega\;.
\end{aligned}
\label{eq:di Balazs}
\end{equation}

For a point dipole oriented parallel to its direction of motion, the radiation is negligible for $\beta \lesssim 1$ \cite{jelley}.

To compute the Cherenkov radiation for finite separations $d$, let us
consider a pair moving in the $+x$ direction.  The pair lies entirely
in the transverse plane $y$-$z$, with the line between them making an
angle $\alpha$ with respect to the $y$-axis.  Then, generalizing
Eq. (\ref{eq:mono current density}), the charge density from the
pair is
\begin{equation*}
\rho(\vec k,\omega) =
\frac{ze}{2\pi}\delta(\omega-k_1v)
\big[e^{-i(k_2y_+ - k_3z_+)} - e^{-i(k_2y_- - k_3z_-)}\big].
\end{equation*}
The two charges have positions relative to the
center of mass
\begin{equation*}
\begin{aligned}
&y_+=\frac{d}{2}\cos\alpha\quad&z_+=-\frac{d}{2}\sin\alpha\\
&y_-=-\frac{d}{2}\cos\alpha&z_-=\frac{d}{2}\sin\alpha\;.
\end{aligned}
\end{equation*}
The angle $\alpha$ is the relative azimuth between the line connecting
the two charges and the azimuth of observation.

The generalization of $E_1(\omega)$ of Eq. \ref{eq:mono E1} is
\begin{align}
E_1(\omega)=&\frac{-ize\omega}{v^2}\sqrt{\frac{\pi}{2}}\left(\frac{1}{\epsilon(\omega)}-\beta^2\right)\nonumber\\
&\times\left[K_0(\lambda b_-)-K_0(\lambda b_+)\right]
\end{align}
where
\begin{equation*}
b_\pm = \sqrt{\frac{d^2}{4}\sin^2\alpha +(b\pm\frac{d}{2}\cos\alpha)^2}\;.
\end{equation*}
As before, we take $|\lambda a| \gg 1$ and $a < b$, so we need only
consider $d \ll b$; there is little interference for $d\gtrsim b$.
Therefore, we can simplify using
\begin{equation*}
b_\pm \simeq b\pm \frac{d}{2}\cos \alpha\;.
\end{equation*}
Then, as before, considering completely imaginary $\lambda$ and
$|\lambda a|\gg 1$,
\begin{equation}
\begin{aligned}
E_1(\omega)=&\frac{2ze\omega}{c^2}\left(1-\frac{1}{\beta^2\epsilon(\omega)}\right)\\
&\times\sqrt{\frac{i}{|\lambda|}}\frac{e^{i|\lambda|b}}{b}\sin\left[\frac{d}{2}|\lambda|\cos\alpha\right]
\end{aligned}
\end{equation}
and a similar expression for $B_3(\omega)$.
Here we have taken $b_-\simeq b_+\simeq b$ in the denominator.  

\begin{figure}[h]
\includegraphics[width=\columnwidth]{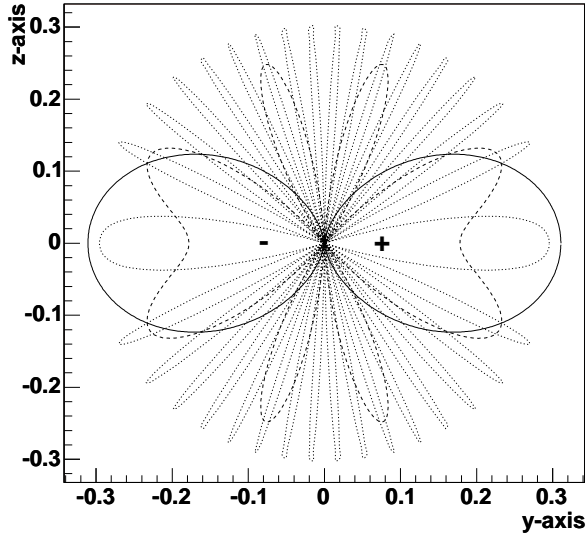}
\caption{The azimuthal angular distribution (transverse to the
direction of motion) of Cherenkov radiation for 500 nm photons from a
pair of charges oriented as shown in the Figure.  Distributions are
shown for pair separations 100 nm (solid line), 1 $\mu$m (dashed line)
and 5 $\mu$m (dotted line), with $\sqrt{\epsilon(\omega)}=n=1.3$ and
$\beta=1$.}
\label{fig:angdist}
\end{figure}

At $\alpha=\pm\pi/2$, $E_1(\omega)=0$.  The Cherenkov radiation is no
longer symmetric about the direction of motion, and vanishes at
right angles to the direction of the dipole. 
As the charge separation increases (or the wavelength decreases), the
angular distribution evolves from two wide lobes into a many-lobed
structure, as shown in Fig. \ref{fig:angdist}.  After integration over
even a narrow range of $\omega$ or $d$, the angular distribution
becomes an almost-complete disk, with two narrow zeroes remaining at a
direction perpendicular to the dipole vector.

\begin{figure}[h]
\includegraphics[width=\columnwidth]{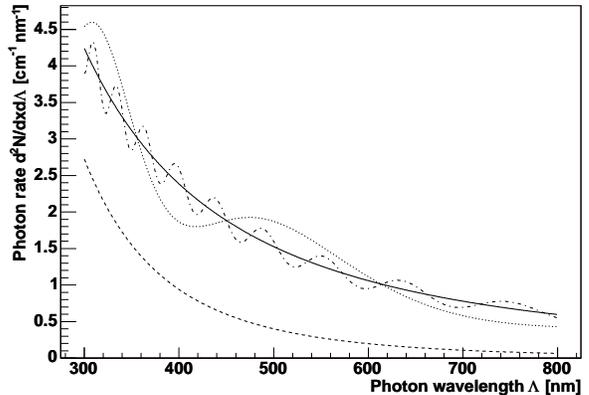}
\caption{The spectrum of Cherenkov radiation at $\beta=1$,
$\sqrt{\epsilon(\omega)}=n=1.3$.  Solid line is for $e^+e^-$ with the
particles considered independently, and the dashed lines are for pairs
treated coherently, with separations 100 nm, 1 $\mu$m and 5 $\mu$m.}
\label{fig:spectrum}
\end{figure}

After assembling the pieces, and averaging over $\alpha$, we find the
generalization of Eq. (\ref{eq:mono dE/dx}),
\begin{equation}
\begin{aligned}
\left(\frac{dE}{dx}\right)=&\frac{(ze)^2}{c^2}\int_{\epsilon(\omega)>1/\beta^2}d\omega\\
&\omega\left(1-\frac{1}{\beta\epsilon(\omega)}\right)
\times 2\left[1-J_0(\lambda d)\right]\;.
\end{aligned}
\label{eq:di dE/dx}
\end{equation}
Here $J_0$ is the zeroth order Bessel function of the first kind.  For
$\lambda d\ll 1$, this reproduces Eq. (\ref{eq:di Balazs}).  For
$\lambda d\gg 1$, the $dE/dx$ is twice that expected for an
independent particle (Eq. (\ref{eq:mono dE/dx})).  The transition is
shown in Fig. \ref{fig:spectrum}.  As the emission wavelength
$\Lambda$ approaches $d$, the pair spectrum converges to the
point-charge spectrum in an oscillatory fashion, characteristic of the
Bessel function.  For certain values of $\lambda d$, the radiation
exceeds that of two independent charged particles.

For the remainder of the paper, we assume that media satisfy
$\sqrt{\epsilon(\omega)}=n$, where $n$ is independent of frequency.
In realistic detection media, any variation of $n$ with frequency is
small, and would have little effect on Cherenkov radiation from
relativistic particles.

With real $e^+e^-$ pairs, two effects should be considered.
Electromagnetic radiation is not emitted instantaneously, but occurs
while the radiating particles travel a distance known as the formation
length, $l_f$.  For Cherenkov radiation, $l_f =
\Lambda/\sin^2(\theta_C) = \Lambda\epsilon\beta^2/(\epsilon\beta^2-1)$
\cite{zolotorev}, depends only on the Cherenkov emission angle,
$\theta_C$ and the photon wavelength; $l_f$ depends only slightly
(through $\beta$) on the electron energy.

While the pair is covering the distance $l_f$, the pair separation
will change by an amount $\Delta d = l_f\sin{(\theta)}$, where
$\theta$ is the angle between the $e^+$ and $e^-$ velocity vectors.
Since  $\theta$ is of order $1/\gamma$,  $\Delta d/d \ll 1$, so
the change in separation is not significant.  

Second, the Cherenkov radiation produced at a point ($x$-coordinate)
depends on the fields emitted by the charged particles at earlier
times, when $d$ may be different than at the point of radiation.  For
full rigor, these retarded separations should be used in the
calculation.  Again, this has a negligible effect on the results.

\section{Radiation from \lowercase{$e^+e^-$} pairs in showers}

Many experiments study Cherenkov radiation from large electromagnetic
showers.  The radiation from a shower may be less than would be
expected if every particle were treated as independent.  We use a
simple simulation to consider 300 to 800 nm radiation from
electromagnetic showers.  This frequency range is typical for
photomultiplier based Cherenkov detectors; at longer wavelength, there
is little radiation, while shorter wavelength light is absorbed by the
glass in the phototube.

We simulated 1000 $\gamma$ conversions to $e^+e^-$ pairs with total
energies from ${10}^8$ to ${10}^{20}$ eV.  Pairs were produced with the
energy partitioned between the $e^+$ and $e^-$ following the
Bethe-Heitler differential cross section $d\sigma\approx
E_\pm(1-E_\pm)$, where $E_\pm$ is the electron (or positron
energy)\cite{may}.  At high energies in dense media (above ${10}^{16}$
eV in water or ice), the LPM effect becomes important, and more asymmetric
pairs predominate \cite{sklpm}.  The pairs are generated with initial
opening angle of $m/k$; the fixed angle is a simplification, but the
pair separation is dominated by multiple scattering, so it has little
effect on our results.

The $e^-$ and $e^+$ are tracked through a water medium (with
$n=\sqrt{\epsilon}=1.3$) in steps of $0.02 X_0$, where $X_0$ is the
radiation length, 36.1 cm in water.  At each step, the particles
multiple-scatter, following a Gaussian approximation
\cite[Ch. 27]{PDG}.  The particles radiate bremsstrahlung photons,
using a simplified model where photon emission follows a Poisson
distribution, with mean free path $X_0$.  Although this model has
almost no soft bremsstrahlung, soft emission has little effect on
Cherenkov radiation, since the electron or positron velocity is only
slightly affected.

At each step, we compute the Cherenkov radiation for each pair.  They
are treated coherently when $d < 2\Lambda$; at larger separations
the particles radiate independently.

\begin{figure}[h]
\includegraphics[width=\columnwidth]{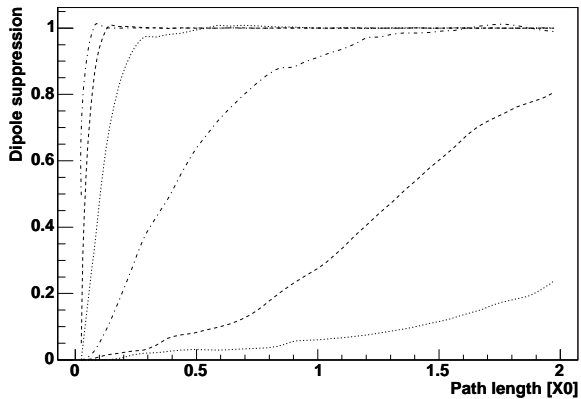}
\caption{Average Cherenkov photon emission rate for pairs with
energies from ${10}^{10}$ (leftmost, dot-dashed curve) to ${10}^{15}$ eV
(rightmost, dotted curve) vs. the distance travelled by the pair in
water, relative to emission from two independent particles.}
\label{fig:ceren}
\end{figure}

As shown in Fig. \ref{fig:ceren}, the particles in lower energy pairs
($< {10}^{10}$ eV) radiate almost independently.  In contrast, the
radiation from very high energy pairs ($> {10}^{15}$ eV) is largely
suppressed.  The broad excursions slightly above unity occur when
$J_0(\lambda d) > 1$ for many of the scattered pairs.

\section{Implications for experiments}

At least two types of astrophysical observatories depend on Cherenkov
radiation.  Water and ice based neutrino observatories observe
Cherenkov radiation from the charged particles produced in neutrino
interactions, and air Cherenkov telescopes look for $\gamma$-ray
induced electromagnetic showers in the Earth's atmosphere.

Current neutrino observatories can search for electron neutrinos with
energies above 50 TeV (for $\nu_\mu$, the threshold is much lower)
\cite{nuobservatories}.  They use large arrays of photomultiplier
tubes to observe the Cherenkov radiation from $\nu_e$ induced showers.
For water, $n\approx 1.3$, Fig. \ref{fig:ceren} shows that $\lambda d <
1$ while the pair travels significant distances. Ice is similar to
water, with a slightly lower density; $n$ of ice depends on its
structure, and is typically $\approx 1.29$ \cite{icen}.

To quantify the effect of Cherenkov radiation from $\nu_e$
interactions, we use a toy model of an electromagnetic shower.  The
shower evolves through generations, with each generation having twice
as many particles as the preceding generation, with half the energy.
Each generation evolves over a distance of $X_0$; other simulations
have evolved generations over a shorter distance $(\ln{2})X_0$,
leading to a more compact shower \cite{showermodel}.  In these
showers, most of the particles are produced in the last radiation
lengths.

\begin{figure}[h]
\includegraphics[width=\columnwidth]{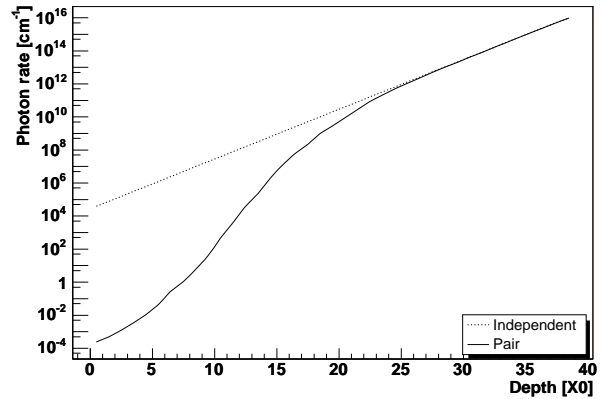}
\caption{Cherenkov radiation from a ${10}^{20}$ eV shower in water,
using the Heitler toy model, versus shower depth (smoothed).  The two
curves compare the radiation for $e^+e^-$ calculated as independent
particles and as coherent pairs.}
\label{fig:toy}
\end{figure}

Fig. \ref{fig:toy} shows the Cherenkov radiation expected from a model
${10}^{20}$ eV shower with coherent Cherenkov radiation (solid line) and
in a model where all particles radiate independently (dotted line).
This model does not include the LPM effect, so it should be considered
only illustrative.  The LPM effect lengthens the high-energy (above a
few ${10}^{15}$ eV) portion of the shower.  By spreading the shower
longitudinally, the LPM effect will give the electrons and positrons
more time to separate, and so will somewhat lessen the difference
between the two results.  However, it is clear from Fig. \ref{fig:toy}
that coherence has a significant effect for the first $\approx 22$
generations.  Since the front of the shower contains relatively few
particles, it will not affect the measured energy; the change in
number of radiated photons (and hence on the energy measurement)
should be less than 1\%.

However, the suppression will affect the apparent length of the shower.
For the first $\approx 8$ generations, the shower will emit less light
than a single charged particle.  Because of the LPM effect, each of
these generations (with mean particle energy $E_g$ above a few
${10}^{15}$ eV) develop over a distance $X = X_0\sqrt{E_g/5E_{LPM}}$,
where $E_{LPM}=278$ TeV for water is the effective LPM energy
\cite{showermodel}, greatly elongating the shower.  So, the first 8
generations include most of the length of the shower.  So, the
suppression of Cherenkov radiation hides the initial shower
development, making the shower appear considerably more compact.  The
reduction in early-stage radiation should help in separating electron
cascades from muon-related backgrounds, especially muons that undergo
hard interactions, and lose a large fraction of their energy.

Atmospheric Cherenkov telescopes like the Whipple observatory study
astrophysical $\gamma$-rays with energies from 100 GeV to 10 TeV.
These telescopes observe Cherenkov radiation from pairs in the upper
atmosphere; for a 1 TeV shower, the maximum particle density occurs at
an altitude of 8 km above sea level (asl) \cite{sinnis}, where the
density is about $1/3$ that at sea level.  Since $n-1$ depends
linearly on the density, at 8 km asl $n-1\approx 1\times {10}^{-4}$, so
for 500 nm photons radiated from ultra-relativistic particles,
$\lambda d < 1$ only for $d < 6\;\mu\mbox{m}$. In this low-density
medium, the effect of the pair opening angle is significant and
multiple scattering is less important.  Pairs with $k<1$ TeV will
separate by 30 $\mu$m in a distance less than 30 meters; at 8 km asl,
this is 3\% of a radiation length.  This distance is too short to
affect the radiation pattern from the shower.

Cherenkov radiation is also used in lead-glass block calorimetry,
and in Cherenkov counters for particle identification; their response
to photon conversions may be affected by this coherence.  

Although the reactions are slightly different, a similar analysis
applies to the reduction of ionization by $e^+e^-$ pairs.  Perkins
observed that the ionization from pairs with mean energy 180 GeV in
emulsion was surppressed for the first $\approx 250$ $\mu$m after the
pairs were created \cite{perkins}.  With $X_0=3$ cm (typical for
emulsion), the $e^+$ and $e^-$ trajectories will be about 4 nm apart
after travelling $250$ $\mu$m.  For relativistic particles, the
screening distance (effective range for $dE/dx$) is determined by the
plasma frequency of the medium, $\omega_p$.  For silver bromide, the
dominant component of emulsion, $\hbar\omega_p = 48$ eV \cite{jackson}
(in a complete emulsion, $\hbar\omega_p$ will be slightly lower). This
yields a screening distance $c/\omega_p = 4$ nm, which is very close
to the calculated separation.

\section{Conclusion}

We have calculated the Cherenkov radiation from $e^+e^-$ pairs
as a function of the pair separation $d$.  When $d^2 < v^2/(\omega^2
[1-\beta^2\epsilon(\omega)])$, the radiation is suppressed
compared to that from two independent particles.  

This suppression affects the radiation from electromagnetic showers in
dense media.  Although the total radiation from a shower is not
affected, emission from the front part of the shower is greatly
reduced; this will affect studies of the shower development, and may
affect measurements of the position of the shower.

This work was funded by the U.S. National Science Foundation
under Grant number OPP-0236449 and the U.S. Department of
Energy under contract number DE-AC-76SF00098.


\end{document}